\documentclass[twocolumn,showpacs,preprintnumbers,amsmath,amssymb]{revtex4}

\usepackage{graphicx}
\usepackage{dcolumn}
\usepackage{bm}
\usepackage{tabularx}

\begin{document}

\title{
Experimental verification of the surface termination in the topological insulator TlBiSe$_{2}$ using core-level photoelectron spectroscopy and scanning tunneling microscopy
}

\author{Kenta~Kuroda$^{1}$, Mao~Ye$^{2,3}$, Eike~F.~Schwier$^{2}$, Munisa~Nurmamat$^{1}$, Kaito~Shirai$^{1}$, Masashi~Nakatake$^{2}$, Shigenori~Ueda$^{4}$, Koji~Miyamoto$^{2}$, Taichi~Okuda$^{2}$, Hirofumi~Namatame$^{2}$, Masaki~Taniguchi$^{1,2}$, Yoshifumi~Ueda$^{5}$, and Akio~Kimura$^{1}$}

\affiliation{
$^{1}$Graduate School of Science, Hiroshima University, 1-3-1 Kagamiyama, Higashi-Hiroshima 739-8526, Japan
 }

\affiliation{
$^{2}$ Hiroshima Synchrotron Radiation Center, Hiroshima University, 2-313 Kagamiyama, Higashi-Hiroshima 739-0046, Japan
}

\affiliation{
$^{3}$ Shanghai Institute of Microsystem and Information Technology, Chinese Academy of Sciences, 865 Chang Ning Road, Shanghai 200050, China
}

\affiliation{
$^{4}$ Synchrotron X-ray Station at SPring-8, National Institute for Materials Science, Hyogo 679-5148, Japan
}

\affiliation{
$^{5}$Kure National College of Technology, Agaminami 2-2-11, Kure 737-8506, Japan
}


\date{\today}

\begin{abstract}
The surface termination of the promising topological insulator TlBiSe$_{2}$ has been studied by surface and bulk sensitive probes. Our scanning tunneling microscopy has unmasked for the first time the unusual surface morphology of TlBiSe$_{2}$ obtained by cleaving, where islands are formed by residual atoms on the cleaved plane. The chemical condition of these islands was identified using core-level spectroscopy. We observed thallium core-level spectra that are strongly deformed by a surface component in sharp contrast to the other elements. We propose a simple explanation for this behavior by assuming that the sample cleaving breaks the bonding between thallium and selenium atoms, leaving the thallium layer partially covering the selenium layer. These findings will assist the interpretation of future experimental and theoretical studies on this surface.
\end{abstract}

\pacs{79.60.Jv, 68.37.Ef, 72.15.Lh, 61.43.-j}

\maketitle
%
\begin{center}
\begin{figure} [t] 
\includegraphics[scale=0.144]{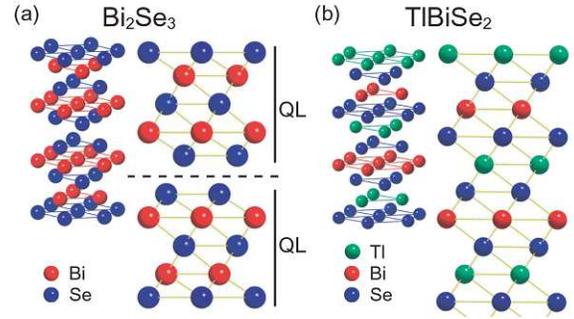}
\caption{(color online) Crystal structures of (a) Bi$_{2}$Se$_{3}$ and (b) TlBiSe$_{2}$. 
(left) Part of the crystal structures where 10 layers are shown. 
(right) Side view from [100] direction. The crystal structure of Bi$_{2}$Se$_{3}$ and 
TlBiSe$_{2}$ with the quintuple layers (QLs) stacking in the sequence Se-Bi-Se-Bi-Se 
and the atomic layer stacking in the sequence -Tl-Se-Bi-Se-Tl- without van-der-Waals-gap.
}
\end{figure}
\end{center}
%
\section{I. Introduction}
Topological insulators are known as a new class of materials where the bulk is an insulator with a band inversion due to a strong spin-orbit coupling, which results in the emergence of topological surface states (TSSs) forming Dirac-cone-like bands with high spin polarizations. Owing to the existence of their non-trivial surface state, the surfaces of topological insulators open the opportunity to realize intriguing electromagnetic properties as well as future spintronic devices or fault-tolerant quantum computers~\cite{Hasan_RMP, Zhang_RMP, Zhang_PRB_08, Zhang_NP_10, Nomura_PRL_11}.

So far, several binary chalcogenides, such as Bi$_{2}$Se$_{3}$ and Bi$_{2}$Te$_{3}$, are known as the prototypical topological insulators. They form a tetradymite crystal structure with a stacking of quintuple layers (QLs) (Se-Bi-Se-Bi-Se atomic layers) weakly coupled via van-der-Waals-forces as shown in Fig. 1(a). Due to the weak bonds between the QLs, the outermost Se layer of a QL terminates the surface after cleaving and it is therefore thought that dangling bonds will not emerge. The scenario that layered materials cleave along their van-der-Waals gap is widely accepted~\cite{Dangling_06} and the chalcogenide termination of the topological insulators has been confirmed by the combination of angle-resolved photoelectron spectroscopy~\cite{Xia_NatPhys_09, Chen_Science_09}, and $ab$-$initio$ calculation~\cite{Zhang_NatPhys_09}. 
 
After finding the tetradymite-type topological insulators, TlBiSe$_{2}$ has been discovered to be a topological insulator~\cite{Sato_PRL_10, Kuroda_PRL_10, Chen_PRL_10}, which forms a three-dimensional crystal structure without van-der-Waals-gaps, as shown in Fig. 1(b). The crystal is built up by stacking [-Tl-Se-Bi-Se-]$^{n}$ layers along the $c$-axis of its hexagonal unit cell. In this compound, the observed TSS features an in-gap Dirac point which is well isolated from the continuum states of the bulk~\cite{Kuroda_PRL_10}. This may prove essential for the realization of an ambipolar gate-control in spin-current devices as well as for studying novel topological properties, where the Dirac point is required to be close to the Fermi energy~\cite{Hasan_RMP, Zhang_RMP, Zhang_PRB_08, Zhang_NP_10, Nomura_PRL_11}. In this respect, TlBiSe$_{2}$ is known as one the most promising topological insulators of today. However, due to the absence of any van-der-Waals-gaps, and the overall covalent and ionic natures of the interatomic bonds, it is still unknown which layer terminates the surface after cleaving. A theoretical study predicted that trivial surface states should co-exist with the non-trivial TSS due to dangling bonds at the cleaved surface and even the TSS will be affected by the surface termination~\cite{Eremeev_PRB_11}. However, no trivial surface states have been observed and the observed band structures of the TSS apparently are the same for previous ARPES measurements~\cite{Kuroda_PRL_10, Sato_PRL_10, Chen_PRL_10}. Therefore, finding the surface termination is a important key to understand the discrepancy between the experimental and theoretical results. 
  
In this paper, we present an experimental approach to determine the surface termination of TlBiSe$_{2}$ by using a combination of scanning tunneling microscopy (STM) and core-level photoelectron spectroscopy (CL-PES). First, we present STM results revealing that residual islands are formed on the cleaved surface. This fact makes a case for a preferential cleaving plane between the Tl and Se layers, leaving a Se surface with the residual Tl atoms forming the islands. This particular situation is strongly supported by employing PES on the core-levels of TlBiSe$_{2}$ using different surface and bulk sensitive photon energies ($h\nu$s) ranging from vacuum ultraviolet (VUV) to the hard x-ray (HAX) regime.
 
\section{II. Experiment}
Single crystalline samples of TlBiSe$_{2}$ were grown by using the Bridgeman method using high purity elements (Bi, Se: 99.999 $\%$, Tl: 99.99 $\%$). The materials were heated in an evacuated quartz ampule above the melting point around 800~$^{\circ}$C, and kept at the constant temperature for two days. It was then cooled down to 100~$^{\circ}$C over a period of twenty days. 

The STM experiment was conducted at 78 K in an ultrahigh vacuum with a base pressure better than 1$\times$10$^{-8}$~Pa using a low temperature scanning tunneling microscope (Omicron Nano Technology). STM images were acquired in the constant-current mode with a bias voltage ($V_{s}$) applied to the sample. The samples were cleaved $in$-$situ$ at room temperature and then transferred to the measurement chamber.

Surface sensitive PES in the VUV regime (VUV-PES) was conducted at BL-7 of the Hiroshima Synchrotron Radiation Center with a hemispherical photoelectron analyzer (VG-SCIENTA SES 2002) at 80~K with ultra-high vacuum conditions better than 1$\times$10$^{-8}$~Pa. The excitation energies ranged from 24 to 250~eV. The total energy resolution for the VUV-PES measurement was obtained by Fermi edge fitting of polycrystalline Au and found to be approximately $E/\Delta E=1000$ . The samples were cleaved $in$-$situ$ at 80 K. The VUV-PES spectra were recorded at normal emission in angle integrated mode.

Bulk sensitive PES with hard x-ray (HAX-PES) was measured at BL15XU of SPring-8~\cite{BL15XU} at room temperature. An excitation energy of $h\nu$=5.95~keV was used, which results in an  inelastic mean free-path (IMFP) for the photoelectrons as large as 50~\AA . This corresponds to a probing depth of up to 30~atomic layers for TlBiSe$_{2}$. The total energy resolution for the HAX-PES measurement was determined to be 246~meV. The sample were cleaved at atmosphere and immediately installed into the vacuum chamber.

%
\begin{center}
\begin{figure} [t] 
\includegraphics[scale=0.144]{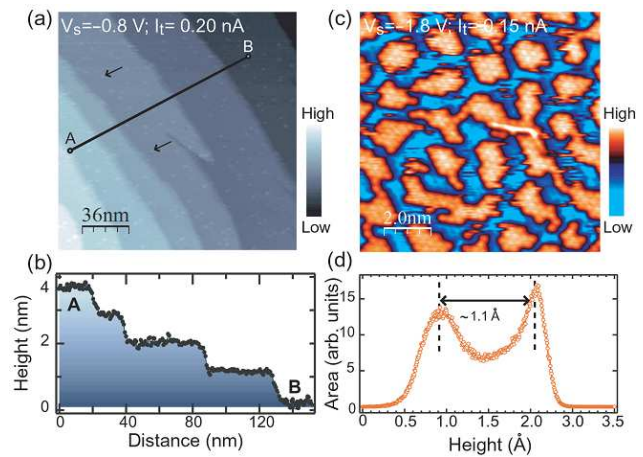}
\caption{(color online) (a) Large scale STM image of TlBiSe$_{2}$; (b) Height profile of the 
steps on the cleaved surface of TlBiSe$_{2}$ along the A-B line in (a); (c) Small area 
of the cleaved surface, showing atomically resolved structure of the residual clusters 
on the cleaved surface; (d) Histogram of the surface shown in (c).}
\end{figure}
\end{center}
%
\section{III. Results and Discussion}
First, the cleaved surface of TlBiSe$_{2}$ has been directly examined by STM. Figure 2(a) shows a typical large area image acquired at a sample bias voltage of $V_{s}$=$-$0.8~V. A couple of step edges are clearly seen in the cleaved plane, whose height profile along A-B is shown in Fig. 2(a). The observed terraces have a typical width larger than 20~nm. It can also be seen that their step height is around 8~\AA , which is comparable to the five-layer-thickness (-Tl-Se-Bi-Se-Tl-). The uniform step height strongly indicates that identical atomic layers terminate the different terraces at the cleaved surface. In figure 2(c), we show a small area STM image on a single terrace with atomic resolution. It can be clearly seen that the terrace is made up by residual islands on top of the surface layer, probably formed by atoms remaining on the cleaved plane. It should be noted that one can see the close-packed lattice structure within the islands, which corresponds to the TlBiSe$_{2}$ (001) surface. These features indicate that the atoms forming the islands remain on top of the surface layer after the sample is cleaved. It is certainly surprising that a single TSS has been identified by ARPES measurement~\cite{Kuroda_PRL_10, Sato_PRL_10, Chen_PRL_10} regardless of the existence of the islands found by the STM. This unambiguously represents the topologically non-trivial character of this surface state.

To qualify the island coverage of the surface and their height distribution, we show that the height distribution of topographic STM image in a histogram in Fig. 2(d), where two peaks can be identified. Note that the height difference between two peaks represents the heights of the island in the topographic image. And the area of the each peak represents the area of islands (right peaks) and the rest area (left peaks), respectively. The height of the islands is found to be 1.1~\AA , which is smaller than the interlayer distance along the $c$-axis $\sim$1.8~\AA ~\cite{Toubektsis_JCG_10}. This difference may indicate the altered local density of states of the islands including an apparent change in height and an inward relaxation of the islands toward the bulk. In both cases, it can be safely concluded that the islands are formed by a single monolayer of atoms. An analysis of the area of the surface and island peaks shows that almost half of the surface is covered by islands. One may notice that bright spots are visible on the terraces (arrows in Fig. 2(a)). By their height, we consider them to be of different surface terminations than the islands. We note however that the occupancy on the terrace is small compared to the islands. Therefore, we can safely disregard their influence on the surface properties.

%
\begin{figure} [t]
\begin{center} 
\includegraphics[scale=0.144]{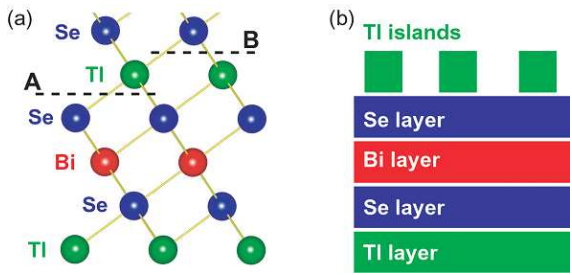}
\caption{(color online)
(a) Schematic image for the cleaving of TlBiSe$_{2}$ from the [100] direction.
Dashed lines with marked A (B) shows the teared bonding with the lower (upper) Se layer. (b) Proposed cleaved surface model.
}
\end{center}
\end{figure}
%
\begin{figure} [t]
\begin{center} 
\includegraphics[scale=0.144]{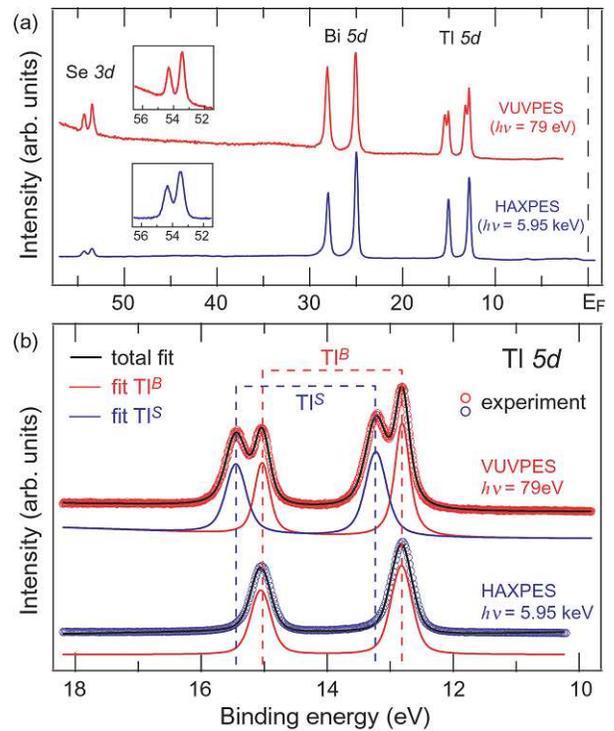}
\caption{(color online)
Results of the photoelectron spectroscopy with vacuum ultraviolet radiation (VUV-PES) 
and hard x-ray (HAX-PES) (a) in the wide energy region and (b) Tl~5$d$ core-level 
energy region. VUV-PES (HAX-PES) result is denoted by upper (lower) line. The solid 
lines in (b) indicate (black) total fitting result, (red) the Tl~5$d$ peaks denoted with 
Tl$^{B}$ as the main line and (blue) Tl$^{S}$ as the satellite line. The observed Tl~5$d$ 
spectra are fitted with Voigt functions and a Shirley-type background. The fitting parameters are listed in Table I.
}
\end{center}
\end{figure}
\begin{table}[t]
\begin{center}
\begin{tabularx}{80mm}{XXX}
\hline
\multicolumn{1}{c}{}&   \multicolumn{1}{c}{Tl$^{B}$}   &   \multicolumn{1}{c}{Tl$^{S}$}\\
\hline \hline
\multicolumn{1}{l}{VUV-PES ($h\nu$=79 eV) } &  & \\
\multicolumn{1}{c}{$E_{B}$ of 5$d_{5/2}$} & \multicolumn{1}{c}{12.81 $\pm$ 0.01} & \multicolumn{1}{c}{13.23 $\pm$ 0.01} \\
\multicolumn{1}{c}{spin-orbit splitting} & \multicolumn{1}{c}{2.22 $\pm$ 0.02} & \multicolumn{1}{c}{2.22 $\pm$ 0.02} \\
\multicolumn{1}{c}{lorentzian width} & \multicolumn{1}{c}{0.16 $\pm$ 0.01} & \multicolumn{1}{c}{0.20 $\pm$ 0.01} \\
\multicolumn{1}{c}{gaussian width} & \multicolumn{1}{c}{0.19 $\pm$ 0.02} & \multicolumn{1}{c}{0.27 $\pm$ 0.02} \\
\multicolumn{1}{c}{$\Delta E$} & \multicolumn{2}{c}{0.42 $\pm$ 0.02} \\
\\
\multicolumn{1}{l}{HAX-PES ($h\nu$=5.95 keV) } &  & \\
\multicolumn{1}{c}{$E_{B}$ of 5$d_{3/2}$} & \multicolumn{1}{c}{12.82 $\pm$ 0.03} & \multicolumn{1}{c}{-} \\
\multicolumn{1}{c}{spin-orbit splitting} & \multicolumn{1}{c}{2.23 $\pm$ 0.05} & \multicolumn{1}{c}{-} \\
\multicolumn{1}{c}{lorentzian width} & \multicolumn{1}{c}{0.16 $\pm$ 0.01} & \multicolumn{1}{c}{-} \\
\multicolumn{1}{c}{gaussian width} & \multicolumn{1}{c}{0.30 $\pm$ 0.02} & \multicolumn{1}{c}{-} \\ 
\hline
\end{tabularx}
\caption{
Fitting parameters for the components of the Tl~$5d$ core-level in Fig. 3(b). Tl$^{B}$ 
and Tl$^{S}$ show the components of the Tl~5$d$ state located at lower and higher $E_{B}$, respectively. 
Listed all energy units are eV. The lorentzian and gaussian widths refer to the full width at half maximum. 
$\Delta E$ indicates the size of splitting for the two components.
}
\end{center}
\end{table}
%
%
%
In the crystal structure of TlBiSe$_{2}$ shown in Fig. 1(b), the Tl layers are sandwiched by Se layers. By taking into account the previous $ab$-$initio$ study, which has shown that the bonding strength between Tl and Se layers is weaker than that for the others~\cite{Eremeev_PRB_11}, it can be expected that cleaving would happen between these layers. Note that there are two possibilities for the surface termination as shown in Fig. 3(a); the breaking of the bond with the lower Se layer (dashed line A in Fig. 3(a)) and the upper layer (dashed line B). In the former case, the Se layer would terminate the surface, whereas the Tl layer will remain on top of the Se layer in the latter case. Since these two possibilities can be considered to be equally likely, we propose a model of the cleaved TlBiSe$_{2}$ surface as seen in Fig. 3(b). The islands on the cleaved surface would then consist of the residual Tl atoms, covering half the Se layer. Note that the Tl-Se swap model where the surface Tl layer is interchanged with the Se layer, i.e. Se-Tl-Bi-Se-Tl-, has been proposed as another possible way to terminate the surface, but the Tl termination was predicted to be energetically preferred to the swap model~\cite{Eremeev_PRB_11}. 

Commonly, a variation of atomic configurations leads to a modification of chemical bonding, which has been widely studied by CL-PES in clusters and step edges at surfaces~\cite{Au_cluster, W_step}. Considering the proposed model as shown in Fig. 3(b), one may expect that the chemical condition of Tl atoms forming the islands and buried in the bulk are different. In this respect, studying the chemical character with CL-PES is useful to examine our model. Therefore, we focus on the core-levels spectra obtained by the surface sensitive VUV-PES and the bulk sensitive HAX-PES techniques. Figure 4(a) shows VUV-PES (top) and HAX-PES (bottom) results in a wide binding energy ($E_{B}$) range. Three core-levels are identified as Se 3$d$, Bi 5$d$ and Tl 5$d$. It becomes evident that VUV-PES results for Bi and Se core-levels are overall similar to those measured by HAX-PES. By contrast, a distinct difference in the spectrum is visible in the Tl 5$d$ core-levels; HAX-PES result shows a single spin-orbit doublet, whereas an additional peak appears for each spin-orbit component in the VUV-PES data. A similar spectral feature of the Tl 5$d$ core-level has already been reported for another thallium based topological insulator TlBiTe$_{2}$ with VUV radiation~\cite{Chen_PRL_10}, however a detailed explanation of this feature was not provided. The magnified Tl 5$d$ core-level spectra are shown in Fig. 4(b), where the red (blue)  circles denote the VUV-PES (HAX-PES) data points. Then, the additional components as seen in the VUV-PES spectra are decomposed by a fitting procedure using four Voigt functions and a Shirley-type background. The fitting results are indicated by the solid lines in Fig. 4(b) and each component is named as Tl$^{B}$ and Tl$^{S}$ for the Tl core-levels located at lower and higher $E_{B}$, respectively. To obtain the best fit, the gaussian width of all Voigt profiles had to be set larger than the experimental energy resolution probably due to lattice vibration~\cite{phonon}. The fitting parameters are tabulated in Table I~\cite{fit_para}. We find that the lorentzian and gaussian widths for Tl$^{S}$ peaks are slightly larger than those for the Tl$^{B}$ peaks. Furthermore, the HAX-PES result shows a single component whose $E_{B}$ corresponds to the Tl$^{B}$ states in the VUV-PES spectra. The size of the splitting between Tl$^{B}$ and Tl$^{S}$ components in the VUV-PES spectrum is estimated to be $\Delta E$=420$\pm$20~meV. Considering the surface sensitivity of VUV-PES experiments as well as the bulk sensitivity of HAX-PES experiments, we identify the Tl$^{S}$ line as a satellite that can be linked to the formation of the islands at the surface as seen in STM and will refer to the Tl$^{B}$ emission as the main line emitted from the bulk.

\begin{figure*}[t]
\begin{center}
\includegraphics[scale=0.144]{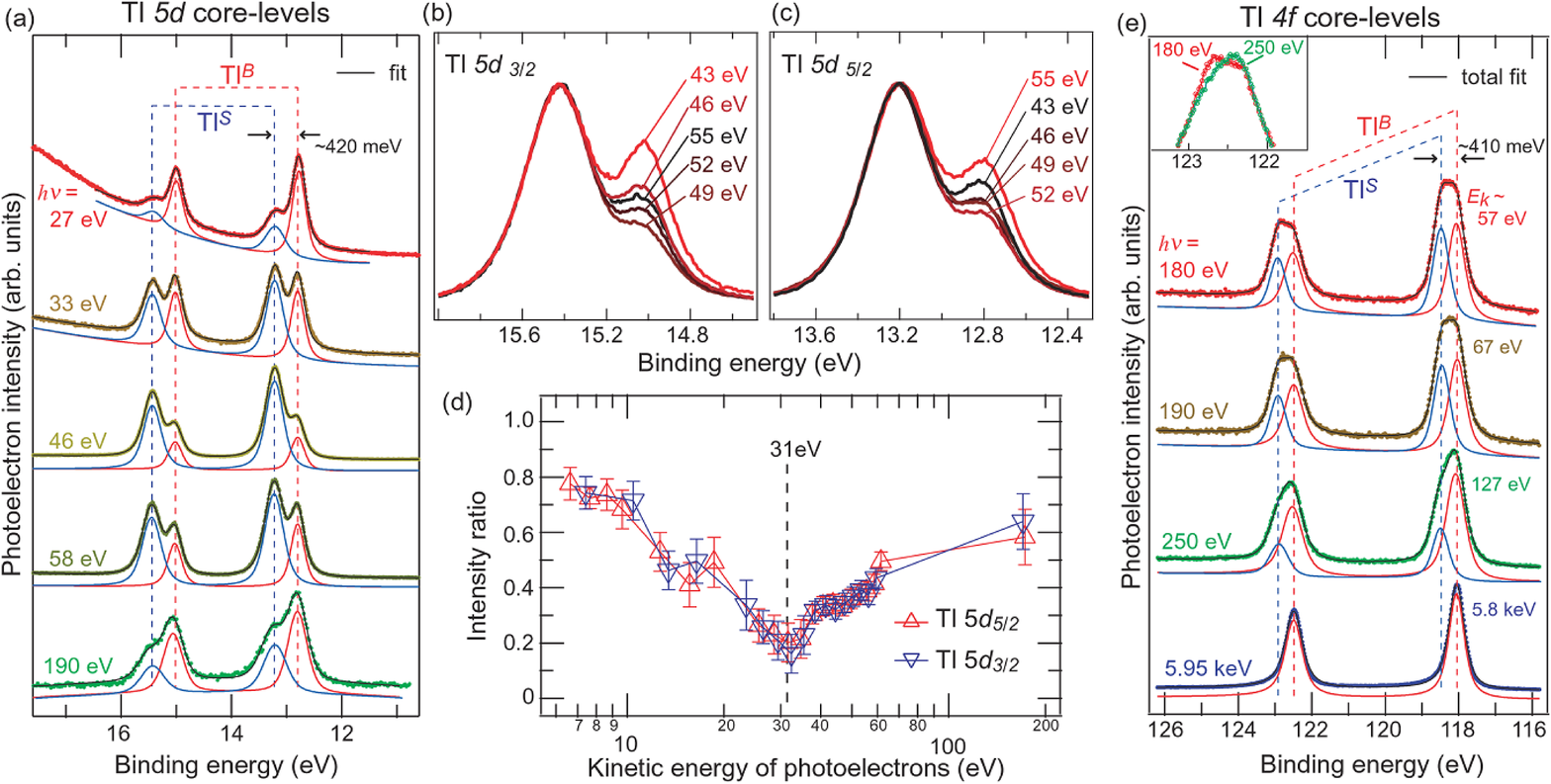}
\caption{(color online)
(a) Core-level spectra for Tl~$5d$ with vacuum ultraviolet (VUV) radiations ranging from 27 to 190~eV. 
The corresponding $E_{kin}$ of the photoelectrons from $5d_{5/2}$ are shown in the figure. 
The red (blue)  solid line shows the Tl$^{B}$ (Tl$^{S}$) lines decomposed with fitting and the 
lorentzian widths are fixed at the same value listed in Table I. (b) and (c) show intensity 
evolution of the Tl$^{B}$ for Tl 5$d_{3/2}$ and Tl 5$d_{5/2}$ with respect to that of Tl$^{S}$ 
peaks, respectively. (d) Intensity ratios between Tl$^{B}$ and Tl$^{S}$ for each doublet 
as a function of $E_{kin}$ of photoelectrons. The intensity ratios are defined with the equation as described 
in the main text. (e) Core-level spectra for Tl~4$f$ with VUV radiations ranging from 
180 to 250~eV as well as hard x-ray.
}
\end{center}
\end{figure*}
For further understanding of this particular spectral feature, we have examined the Tl 5$d$ core-level spectra using various excitation energies. The measurement mirrors similar experiments used to separate surface and bulk components at conventional semiconductor surfaces~\cite{Himpsel_PRL_80, Uhrberg_PRB_94}. In these reports, the kinetic energy ($E_{kin}$) of photoelectrons was tuned to a low energy below 10~eV, where the IMFP increases towards $\sim$10~\AA . This is larger than the five-layer-thickness of TlBiSe$_{2}$ and can be expected to result in a mixture between surface and bulk sensitivity. To continuously tune the IMFP, we use the tunable photon source provided by the synchrotron radiation in the $h\nu$ range of 24-190~eV. The observed spectra are summarized in the Fig. 5(a) together with their fitting results. The lorentzian widths shown in Table I are used for the fitting functions. The total spectra are found to strongly depend on $h\nu$. At a more bulk sensitive excitation energy ($h\nu$=27~eV), the main Tl 5$d$ core-level emission (Tl$^{B}$) is stronger than the emission from the satellite (Tl$^{S}$). Increasing $h\nu$ to tune the $E_{kin}$ towards more surface sensitive levels, the intensity of Tl$^{S}$ becomes stronger than the main line at $h\nu$=46~eV. Further increasing the excitation energy towards $h\nu$=190~eV again reduces the weight of Tl$^{S}$ close to its low $h\nu$ value. This particular $h\nu$ dependences can be reasonably explained by assuming a change of the surface sensitivity and will be discussed further below.

Figures 5(b) and (c) show a magnified view of the Tl 5$d$ core-level spectra normalized by the intensity of the Tl$^{S}$ peak in the energy window of Tl 5$d_{3/2}$ and Tl 5$d_{5/2}$ components, respectively. For both peaks in the doublet, a similar $h\nu$ dependence is observed, except for the fact that the intensity of Tl$^{B}$ for the 5$d_{3/2}$ and 5$d_{5/2}$ peaks goes through a minimum at $h\nu$=52~eV (5$d_{3/2}$) instead of 49~eV (5$d_{5/2}$). This corresponds roughly to the spin-orbit splitting of the doublet and demonstrates that the changes in the intensity of the spectra can be better explained with a dependence on $E_{kin}$ rather than $h\nu$. In the following, we will analyze the intensity ratio between main line and its satellite by defining the normalized intensity ratio: 
\begin{equation}
r_{j} = \frac{I^{B}_{j}}{I^{S}_{j} +I^{B}_{j}}\ 
\end{equation}
where $I^{B}_{j}$ is the fitted intensity of the Tl$^{B}$ line and $I^{S}_{j}$ the corresponding intensity from the Tl$^{S}$ line of the 5$d_{j}$, respectively ($j$=$5/2$ or $3/2$). The $E_{kin}$ dependence of $r_{5/2}$ and $r_{3/2}$ are summarized in Fig. 5(d). As suggested from the preceding qualitative analysis in Fig. 5(b) and (c), it is found that the $E_{kin}$ dependences of both Tl 5$d$ doublets are indeed identical. The intensities of the Tl$^{B}$ lines increase towards lower and higher $E_{kin}$ from their minimum near $E_{kin}$=30~eV. The universal curve of photoelectron IMFP takes the smallest value around $E_{kin}$ of 30-50~eV~\cite{IMFP} and therefore the enhancement of the Tl$^{S}$ state near $E_{kin}$=30~eV indicates that Tl$^{S}$ and Tl$^{B}$ states originate from the surface and bulk, respectively. 

To analyze the relation between Tl emission from the surface and the bulk, we focus on the core-level shift (CLS) between the Tl$^{S}$ and Tl$^{B}$ states. The size of the CLS generally depends on several parameters, such as the bonding distance, the valency of the atom~\cite{Chemical} and the local chemical surrounding. The Tl 5$d$ core-level is normally thought as a semicore state because it is located energetically close to the valence band ($E_{B}$=12-15~eV) and thus it is highly sensitive to the physical and chemical environments compared to other deeper lying core-levels. Therefore, in order to reduce the possibility that shifts are induced from beyond the nearest neighbor lattice site, it is important to compare the CLS of the Tl 5$d$ line with other Tl core-levels that are more strongly screened. To do so, we chose to investigate the Tl 4$f$ core-level, which is located at higher $E_{B}$ ($\sim$118~eV) than that of the Se~3$d$, Bi~5$d$ and Tl~5$d$ core-levels. Figure 4(e) shows the spectral features for Tl~4$f$ core-levels excited with several different $h\nu$ and fitted with Voigt profiles. Even without resorting to an analysis of the fitting results, one can already see that the spectral weight on the higher $E_{B}$ side increases, if the $E_{kin}$ of photoelectrons is close to the highly surface sensitive values (see the inset). Accordingly, the spectral weight at lower $E_{B}$ increases at more bulk sensitive $E_{kin}$. This behavior is verified in more detail by the fitting functions. Similar to Tl~5$d$, we are able to identify a Tl$^{B}$ main line by comparing the VUV-PES data with the spectrum taken with HAX-PES which again shows a single component. The Tl$^{B}$ and Tl$^{S}$ lines of the Tl~4$f$ peak show comparable $E_{kin}$ dependence of the Tl~5$d$ core-levels. The size of the CLS between these lines is estimated to be 410$\pm$20~meV which is well comparable to that of the Tl~5$d$ level. This finding again support our notion that the chemical state of Tl is strongly deformed by the chemical environment at the surface.

\begin{figure} [t] 
\begin{center}
\includegraphics[scale=0.144]{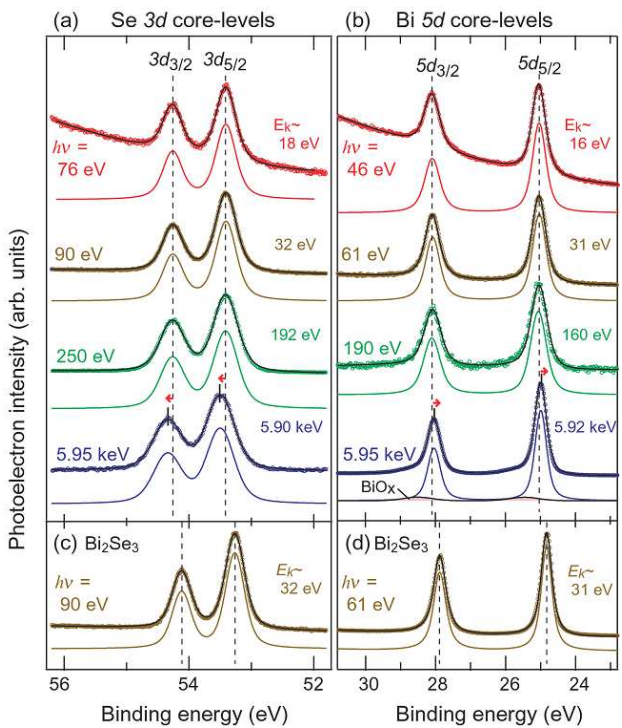} 
\caption{(color online)
Core-level spectra for (a) Se $3d$ and (b) Bi $5d$ with vacuum ultraviolet radiations and hard x-ray. 
The corresponding $E_{kin}$ of the photoelectron from $3d_{5/2}$ and $5d_{5/2}$ are denoted 
in the figure. Solid lines show fitting results with parameters listed in Table~II. 
We found the tails in Bi $5d$ spectra 
taken with hard x-ray (bottom) at higher $E_{B}$ which is attributed to the oxidized Bi (BiO$_{x}$)
(shaded area). (c) and (d) are Se $3d$ and Bi $5d$ core-level spectra for Bi$_{2}$Se$_{3}$ obtained 
by the selected photon energy to obtain the most surface sensitivity. 
}
\end{center}
\end{figure}
\begin{table}[t]
\begin{center}
\begin{tabularx}{80mm}{XXX}
\hline
 & \multicolumn{1}{c}{TlBiSe$_{2}$} & \multicolumn{1}{c}{Bi$_{2}$Se$_{3}$} \\
\hline \hline
\multicolumn{1}{l}{Bi $5d$ core-level} & & \\
\multicolumn{1}{l}{$E_{B}$ of $5d_{5/2}$} & \multicolumn{1}{c}{24.98 $\pm$0.01} & \multicolumn{1}{c}{24.83 $\pm$0.01} \\
& \multicolumn{1}{c}{(25.06 $\pm$0.03)} &  \\ 
\multicolumn{1}{l}{spin-orbit split} & \multicolumn{1}{c}{3.04 $\pm$0.02} & \multicolumn{1}{c}{3.05 $\pm$0.02} \\
& \multicolumn{1}{c}{(3.03 $\pm$ 0.05)} &  \\ 
\multicolumn{1}{l}{lorentzian width of $5d_{5/2}$} & \multicolumn{1}{c}{0.32 $\pm$ 0.02} & \multicolumn{1}{c}{0.28 $\pm$ 0.02} \\
& \multicolumn{1}{c}{(0.20 $\pm$ 0.02)} &  \\ 
\multicolumn{1}{l}{lorentzian width of $5d_{3/2}$} & \multicolumn{1}{c}{0.38 $\pm$ 0.02} & \multicolumn{1}{c}{0.34 $\pm$ 0.02} \\
& \multicolumn{1}{c}{(0.24 $\pm$ 0.02)} &  \\ 
\\
\multicolumn{1}{l}{Se $3d$ core-level} & & \\
\multicolumn{1}{l}{$E_{B}$ of $3d_{5/2}$} & \multicolumn{1}{c}{53.50 $\pm$ 0.01} & \multicolumn{1}{c}{53.26 $\pm$ 0.01} \\ 
& \multicolumn{1}{c}{(53.41 $\pm$ 0.03)} &  \\  
\multicolumn{1}{l}{spin-orbit split} & \multicolumn{1}{c}{0.84 $\pm$ 0.02} & \multicolumn{1}{c}{0.86 $\pm$ 0.02} \\
& \multicolumn{1}{c}{(0.85 $\pm$ 0.05)} &  \\ 
\multicolumn{1}{l}{lorentzian width of $3d_{5/2}$} & \multicolumn{1}{c}{0.20 $\pm$ 0.01} & \multicolumn{1}{c}{0.16 $\pm$ 0.01} \\
& \multicolumn{1}{c}{(0.20 $\pm$ 0.01)} &  \\ 
\multicolumn{1}{l}{lorentzian width of $3d_{3/2}$} & \multicolumn{1}{c}{0.20 $\pm$ 0.01} & \multicolumn{1}{c}{0.16 $\pm$ 0.01} \\
& \multicolumn{1}{c}{(0.20 $\pm$ 0.01)} &  \\ 
\hline
\end{tabularx}
\caption{
Fitting parameters of Bi $5d$ and Se $3d$ spectra in TlBiSe$_{2}$ and Bi$_{2}$Se$_{3}$. 
The values of HAX-PES are shown in parentheses. The lorentzian width refer to the full 
width at half maximum~\cite{Gauss}. All listed energy units are in eV. 
}
\end{center}
\end{table}
%
In order to get a proof that no similar CLS exists for the Se or Bi atoms, we focus on the shallow core-levels from these elements next. Figures 6(a) and (b) summarize the Se~3$d$ and Bi~5$d$ core-level spectra, respectively. The fitting parameters are as listed in Table II and compared with those for Bi$_{2}$Se$_{3}$ as a reference (as discussed later). We find that no additional features, such as shoulders or peak shifts are present in the VUV-PES spectra. Both the Se~3$d$ and the Bi~5$d$ core-level spectra can be reproduced by a single Voigt profiles with the same parameters listed in Table II~\cite{Gauss}. Tuning $h\nu$ to obtain a highly surface sensitive condition ($E_{kin}\sim$30~eV) does not change the spectral shape, which is in a strong contrast to the Tl core-levels. We notice that the peak positions slightly shift in different directions for the Se~3$d$ and Bi~5$d$ peaks if the HAX-PES results are compared to those from the VUV-PES measurements. The Bi~5$d$ spectrum measured by HAX-PES is located at lower $E_{B}$ with respect to that acquired by the VUV-PES while the Se~3$d$ shifts to higher $E_{B}$ with a comparable energy shift of $\sim$80~meV. In addition, the lorentzian width of the HAX-PES peak for Bi~5$d$ is smaller than that measured by VUV-PES. This result probably indicates that all observed VUV-PES spectra of Bi~5$d$ include an unresolved surface component.  

In order to understand the possible contribution of dangling bonds on the Bi~5$d$ and Se~3$d$ core-levels, we investigate the core-levels of Bi$_{2}$Se$_{3}$ in which dangling bonds are believed to be absent.  Figures 6(c) and (d) show the observed Se~3$d$ and Bi~5$d$ spectra in Bi$_{2}$Se$_{3}$ at highly surface sensitive $E_{kin}$. We find that both core-levels are located at lower $E_{B}$ with respect to those of TlBiSe$_{2}$, indicating the different bonding conditions as well as spontaneous carrier doping effect due to the presence of defects in the bulk for Bi$_{2}$Se$_{3}$. We note that the spectral shape in Bi$_{2}$Se$_{3}$ is quantitatively close to that in TlBiSe$_{2}$ for both core-levels and these features are independent of the $E_{kin}$ (not shown). Thus, the contribution of the dangling bonds is found to be weak for Bi and Se core-levels in TlBiSe$_{2}$ and we consider that an unresolved surface component in TlBiSe$_{2}$ may be attributed to a surface relaxation effect which is predicted to decay slowly into the bulk~\cite{Eremeev_PRB_11}. A previous CL-PES measurement on Bi$_{2}$Se$_{3}$ with Fe deposition has demonstrated that the spectral feature of Se 3$d$ core-level are stable against the chemical environment~\cite{Rader_Fe}. Hence, we do not deny the existence of the Se atoms affected by the different chemical surroundings in TlBiSe$_{2}$.

The proposed model for the surface termination as shown in Fig. 3(b) is therefore supported by our core-level PES where only the Tl core-levels are strongly deformed by the surface component. Since a flat surface is generally assumed for the previous calculation~\cite{Eremeev_PRB_11}, the fact that residual islands exist on the surface provides us with a reason why the trivial surface states which present in the calculated band structure are absent in the ARPES measurements~\cite{Sato_PRL_10, Kuroda_PRL_10, Chen_PRL_10}. In the following, we want to give three possible explanations based on our model. Firstly, the non-periodic feature of the residuals islands can result in the absence of dispersing features~\cite{LEED}, as dangling bond states would be localized on the islands as well as in the Se terminated surface area. The localized states may be energetically located above the Fermi energy or weakly contributing to the intensity of photoelectrons and thus be absent from the ARPES measurements. Secondly as mentioned in the STM analysis of the island height the smaller lattice constant than in the bulk may lead to a saturation of the dangling bonds. Finally, a self-doping effect between the dangling bonds of the residual Tl and the Se layers may saturate and the trivial surface states. The presented model may motivate future experiments on this surface. Possible experiments include the study of surface deposition which has been widely used on the surface of Bi$_{2}$Se$_{3}$~\cite{Rader_Fe, Bianchi_BB, Wray_Fe, Ye_Co, Valla_Gd}, as well as theoretical studies on the surface of TlBiSe$_{2}$ including the island structure.
\section{IV. conclusion}
In conclusion, we have experimentally approached the open question of the surface termination problem of TlBiSe$_{2}$ utilizing STM and CL-PES. Our STM results have shown that islands are formed on the cleaved surface. Our CL-PES measurements with tunable surface sensitivity have revealed that the spectral features of Tl core-levels are strongly deformed by the surface component in contrast to the other elements. From these findings, we proposed a model for the surface termination that includes Tl islands covering half of the surface which is terminated by a Se layer. We demonstrated that the residual-Tl-island-model is the most likely explanation for the pronounced CLS found in Tl core-levels. The model was also able to account for the absence of the trivial surface state in ARPES measurements. It will motivate further experimental and theoretical studies of the surface properties.

\section{acknowledgments}
We thank K. Mimura and H. Sato for fruitful discussion. The STM and VUV-PES experiments were performed at HiSOR, 
Hiroshima University with the approval of the Proposal Assessing Committee of HSRC (Proposal No. 12-B-46, 11-B-12). 
The HAXPES experiment was performed at BL15XU of SPring-8 with the approval of NIMS Beamline Station 
(Proposal No. 2012B4908). This work was partly supported by KAK-ENHI (20340092), Grant-in-Aid for Scientific Research 
(B) of Japan Society for the Promotion of Science.


\begin{thebibliography}{12}
\bibitem{fit_para} Lorentzian widths and size of spin-orbit splittings were determined by averaged the all fitting results for each photon energy. 
\bibitem{Gauss} Gaussian widths obtained by the fitting procedure for Tl $5d$ core-level was used for Bi $5d$ and Se $3d$.
\bibitem{LEED} We checked surface diffraction pattern in the cleaved surface of TlBiSe$_{2}$ with low-energy electron diffraction at low temperature $\sim$30~K but only (1$\times$1) are visible and there are no more noteworthy features. Thus, it can be concluded that the islands represent no periodic feature.
\end{thebibliography}
\end{document}